\documentclass[prd,nofootinbib,preprint,superscriptaddress%,showpacs
]{revtex4}

\usepackage{graphicx}
\usepackage{amsfonts}
\newcommand{\be}{\begin{equation}}
\newcommand{\ee}{\end{equation}}
\newcommand{\bea}{\begin{eqnarray}}
\newcommand{\eea}{\end{eqnarray}}
\newcommand{\beq}{\begin{equation}}
\newcommand{\eeq}{\end{equation}}
\newcommand{\beqa}{\begin{eqnarray}}
\newcommand{\eeqa}{\end{eqnarray}}
\newcommand{\no}{\nonumber}
\def\lsim{\lesssim}
\def\gsim{\gtrsim}

\begin{document}
%\preprint{{\vbox{\hbox{}\hbox{}\hbox{}
\hbox{DO-TH-08/02}
\hbox{WIS/04/08-Feb-DPP}
%\hbox{hep-ph/yymmnnn}}}}

\vspace*{.0cm}

\title{Measuring Flavor Mixing with Minimal Flavor Violation\\ at the LHC}

\author{Gudrun Hiller}\email{ghiller@physik.uni-dortmund.de}
\affiliation{Institut f\"ur Physik, Technische Universit\"at Dortmund,
  D-44221 Dortmund, Germany}
\author{Yosef Nir\footnote{The Amos de-Shalit chair of theoretical
    physics}}\email{yosef.nir@weizmann.ac.il}
\affiliation{Department of Particle Physics,
  Weizmann Institute of Science, Rehovot 76100, Israel\vspace*{1cm}}

\vspace*{1cm}
\begin{abstract}
The mixing between third and second (or first) generation squarks is
very small in supersymmetric models with minimal flavor violation such
as gauge-, anomaly- or gaugino-mediation. An opportunity to
measure this mixing will arise if the lightest stop is close enough in
mass to the lightest neutralino, so that the decays into
third generation quarks are kinematically forbidden. We analyze under
which circumstances it might become possible to measure at the
Large Hadron Collider (LHC) the rate of the flavor changing stop decays.

\end{abstract}

\maketitle

%%%%%%%%%%%%%%%%%%%%%
\section{Introduction}
\label{sec:intro}
Significant progress has been achieved in recent years in flavor
precision measurements. All measurements are, however, consistent with
the Standard Model picture, whereby the only source of violation of
the global $SU(3)^5$ symmetry of the gauge interactions are the quark
and lepton Yukawa interactions. Such a situation is not expected if
there is new physics at the TeV scale with generic flavor structure
(for a review, see Ref. \cite{Nir:2007xn}).  This so-called ``New
Physics Flavor Puzzle'' is, however, solved if the new physics is
subject to the principle of Minimal Flavor Violation (MFV)
\cite{D'Ambrosio:2002ex}. This principle states that the Yukawa
interactions remain the only source of the $SU(3)^5$ breaking even in
the presence of new physics.  Known examples of models in this class
are supersymmetric models with gauge-, anomaly-, or gaugino-mediation
of supersymmetry breaking.  More concretely, MFV in the quark sector
implies  that the only spurions that break the global flavor symmetry
\beq
G_q=SU(3)_{Q_L}\times SU(3)_{U_R}\times SU(3)_{D_R}
\eeq
are the up- and down-Yukawa matrices, with the following
transformation properties under $G_q$:
\beq
Y_u(3,\bar3,1),\ \ \ \ Y_d(3,1,\bar3).
\eeq

One of the most powerful predictions of MFV for new physics models is
that flavor mixing is always proportional to the off-diagonal CKM
elements.  Since the elements that connect the third generation to the
two lighter ones, $V_{ub},V_{cb},V_{td}$ and $V_{ts}$, have magnitudes
in the range $0.004-0.04$, the third generation is almost decoupled
from the first two \cite{Grossman:2007bd}.  In the context of
supersymmetry, this aspect of MFV has the following consequences: The
third generation squarks, $\tilde t_L,\tilde b_L, \tilde t_R$ and
$\tilde b_R$, decay predominantly into the third generation quarks, $t$
and $b$; The branching ratios into the lighter generations are
$\lsim|V_{ts}|^2\sim2\times10^{-3}$.  If MFV is strongly violated, it
may be possible to experimentally exclude it by observing decays into
the lighter generations with branching ratios that are significantly
larger than that. (For indirect tests, see, e.g., \cite{Hiller:2003di}.)
If, however, MFV applies, it will be a much more
challenging task to establish it.  One would like to measure the
sub-dominant branching ratios and show that they indeed have the size
predicted by the CKM suppression.  But tagging flavor with such an
accuracy is probably beyond the capabilities of the ATLAS and CMS experiments.

Here we point out that, under certain circumstances, measuring the
decay {\it rate} (rather than the {\it branching ratio}) of a third
generation squark into non-third generation quarks might become
possible. What is required is an approximate degeneracy between the
lightest stop and the lightest neutralino. Then, if the
higgsino/gaugino decomposition of the neutralino, and
the left/right decomposition of the stop are known, one can factor out
the flavor suppression in the decay, and ask whether it fits the CKM
dependence predicted by MFV, or not.

Our main emphasis in measuring intergenerational squark mixing will be 
on flavor changing neutral current (FCNC) decays. Charged current 
processes have always a tree level component inherited 
from the Standard Model, irrespective of the flavor structure of the
supersymmetry breaking. 

The plan of the paper is as follows: In Section \ref{sec:decrat} we
give the ingredients for the stop to have a picosecond lifetime
dominated by FCNC decays. The magnitude of the relevant stop couplings
in MFV are worked out in Section \ref{sec:mfv}.  Prospects for the LHC
making stop decay length measurements are analyzed in Section
\ref{sec:lhc}. We discuss the backgrounds from stop four-body decays
in Section \ref{sec:4body}.  We further comment in Section
\ref{sec:chargino} on a variant of our scenario with a light chargino,
where the stop decays predominantly through charged current
interactions. 
In Section \ref{sec:nonmfv} we analyze models with
quark-squark alignment, which provide an alternative solution
to the flavor puzzle without MFV,
and argue that they can differ significantly in their predictions for
the relevant flavor changing couplings from MFV models. We conclude
in Section \ref{sec:con}.

%%%%%%%%%%%%
\section{The $\tilde t\to c\chi^0$ Decay Rate}
\label{sec:decrat}
Consider a situation where the light stop is the next-to-lightest
supersymmetric particle (NLSP) and that,
furthermore, its decays into final third generation quarks are
kinematically forbidden:
\beq\label{degen}
m_{\tilde t_1}-m_{\chi^0_1}\leq m_b.
\eeq
(From here on, we consider only the lightest stop and the lightest
neutralino, and omit the sub-index 1.) Then, with MFV, the leading
decay mode into second generation quarks is:
\beq\label{domdec}
\tilde t\to\chi^0+c.
\eeq
The question of interest to us is whether such fortunate circumstances
can be exploited to measure the decay rate, with the goal of examining
whether it is suppressed (or not) as predicted by MFV.

Let us denote the $\tilde t-c-\chi^0$ coupling by $Y$. More specifically,
$Y=\sqrt{|y_L|^2+|y_R|^2}$, where the FCNC couplings $y_L,y_R$ parameterize
\beq \label{eq:L}
{\cal{L}}_{c \tilde t \chi^0} =\bar c (y_L P_L +y_R P_R) \chi^0 \tilde t
+ {\rm h.c.},
\eeq
with the chiral projectors $P_{L/R}=(1 \mp \gamma_5)/2$. We further
define $M\equiv m_{\tilde t}$, $m\equiv m_{\chi^0}$ and $\Delta m\equiv
M-m$. We approximate in kinematical factors $m_c\approx0$ and, as
explained above, consider the case $\Delta m\ll M$. The
$\tilde t \to c \chi^0$ decay rate is then given by
\beq\label{decrat}
\Gamma=\frac{M Y^2}{16\pi}\left(1-\frac{m^2}{M^2}\right)^2
\approx\frac{M Y^2}{4\pi}\left(\frac{\Delta m}{M}\right)^2.
\eeq
To get a rough idea of the flavor suppression in $Y$ required for a
long living stop, we rewrite Eq.~(\ref{decrat}) as follows:
\beq\label{lifetime}
\tau_{\tilde t}\sim\ {\rm ps}\ \left(\frac{M}{100\ \mbox{GeV}}\right)
\left(\frac{0.03}{\Delta m/M}\right)^2\left(\frac{10^{-5}}{Y}\right)^2.
\eeq
With such a large lifetime, $\tau_{\tilde t} \gg 1/ \Lambda_{QCD} \sim
10^{-24}$ s, the light stop hadronizes before decay.

Our scenario contains phase space suppression by construction to avoid
stop decays to $t$ or $b$ quarks. We assume $\Delta m/M={\cal
  O}(0.03)$ unless otherwise stated. Within the MFV framework, the
value of $Y$ is driven by the quark Yukawa couplings, and depends on
the $\tilde t_L-\tilde t_R$ decomposition of the $\tilde t$ and the
$\tilde H^0-\tilde w^0-\tilde B$ decomposition of the $\chi^0$. We
work out the size of $Y$ within MFV in the next section.

%%%%%%%%%%%%%%%
\section{Third generation flavor mixing with MFV}
\label{sec:mfv}
Following the MFV rules, we can write the relevant supersymmetry
breaking squark mass terms up to higher powers of the
quark Yukawa couplings  \cite{D'Ambrosio:2002ex}:
\beqa
\tilde m^2_{Q_L}&=&\tilde m^2(a_1{\bf 1}+b_1
Y_uY_u^\dagger+b_2Y_dY_d^\dagger),\no\\
\tilde m^2_{U_R}&=&\tilde m^2(a_2{\bf 1}+b_5 Y_u^\dagger Y_u
+c_1Y_u^\dagger Y_dY_d^\dagger Y_u),\no\\
A_u&=&A(a_4{\bf 1}+b_7 Y_dY_d^\dagger)Y_u.
\label{eq:M23}
\eeqa
We use here the notation of Ref. \cite{D'Ambrosio:2002ex}. We omit
their $b_{3,4}$ terms that are not important for our purposes,
and add the $c_{1}$ term that, albeit small, can be important
in our context. Since we are interested in the $\tilde t_{L,R}$ couplings,
it is convenient to work in the up mass basis, that is,
\beq
Y_u=\lambda_u,\ \ \ Y_d=V\lambda_d,
\eeq
where $V$ denotes the CKM matrix  and $\lambda_q$ are the generation
diagonal Yukawa matrices. We are particularly interested
in the $2-3$ elements of the up squark mass matrices:
\beqa
(\tilde m^2_{Q_L})_{23}&=&\tilde m^2 b_2\lambda_b^2V_{cb}V_{tb}^*,\no\\
(\tilde m^2_{U_R})_{23}&=&\tilde m^2 c_1\lambda_c\lambda_t
\lambda_b^2V_{cb}V_{tb}^*,\no\\
(A_u)_{23}&=&Ab_7\lambda_t\lambda_b^2V_{cb}V_{tb}^*,\no\\
(A_u)_{32}&=&Ab_7\lambda_c\lambda_b^2V_{cb}^*V_{tb}.
\label{eq:MFV23}
\eeqa
Stop-scharm mixing in MFV thus requires at least two powers of the bottom
Yukawa and is CKM suppressed by $V_{cb}$.

We separate the small flavor mixing effects, which we treat as mass
insertions, from possible large mixings, unsuppressed by flavor. In
particular, we take the decomposition of the
light stop mass eigenstate as follows:
\beq
\tilde t (\equiv \tilde t_1)=\cos\theta_{\tilde t}\tilde t_R
+\sin\theta_{\tilde t}\tilde t_L,
\eeq
and explore the full range for $|\sin\theta_{\tilde t}|$ (between 0
and 1).

Similarly, we consider an arbitrary decomposition of the
light neutralino,
\beq
\chi^0 (\equiv \chi^0_1) =V_{1B}\tilde B+V_{1w}\tilde w^0
+V_{1u}\tilde H_u^0+V_{1d}\tilde H_d^{0}.
\eeq

{}From tree-level, single squark mass insertions to the
$\tilde c \tilde B c$, $\tilde c \tilde w^0 c$ and $\tilde c \tilde
H_u^0 c$ vertex (for supersymmetric Feynman rules see, e.g.,
\cite{Rosiek:1995kg}) one finds:
\begin{itemize}
\item The $\tilde t_R-c_R-\tilde B$ coupling is induced by
$(\tilde m^2_{U_R})_{23}$. The
 $\tilde t_R-c_R-\tilde w^0$ coupling vanishes.
\item The $\tilde t_R-c_L-(\tilde w^0,\tilde B)$ coupling is induced by
$(A_u)_{23}$.
\item The $\tilde t_R-c_R-\tilde H_u^0$ coupling is induced by
a combination of $(A_u)_{23}$ and  $\lambda_c$.
\item The $\tilde t_R-c_L-\tilde H_u^0$ coupling is induced by
a combination of $(\tilde m^2_{U_R})_{23}$ and  $\lambda_c$.
\item The $\tilde t_L-c_L-(\tilde w^0,\tilde B)$ coupling is determined by
$(\tilde  m^2_{Q_L})_{23}$.
\item The $\tilde t_L-c_R-\tilde B$ coupling is induced by
$(A_u)_{32}$. The
 $\tilde t_L-c_R-\tilde w^0$ coupling vanishes.
\item The $\tilde t_L-c_L-\tilde H_u^0$ coupling is induced by
a combination of $\lambda_c$ and $(A_u)_{32}$.
\item The $\tilde t_L-c_R-\tilde H_u^0$ coupling is induced by
a combination of $\lambda_c$ and $(\tilde m^2_{Q_L})_{23}$.
\end{itemize}

The situation is summarized in Table \ref{tab:couplings}, where the
gauge and flavor factors and the numerical size of $Y$ modulo the MFV
coefficients $b_i$ defined in  Eq.~(\ref{eq:MFV23})
are  given for the various cases. We use $|V_{cb}V_{tb}^*|\sim0.04$,
$\lambda_t\sim1$,
$\lambda_b^2\sim10^{-3}\tan^2\beta$, and $\lambda_c\sim10^{-2}$ and denote by
$I_3=1/2$ and $Y_Q=1/6$ the weak isospin and
hypercharge of the charm (s)quark doublets.

\begin{table}[t]
\caption{
Flavor structure and numerical size of the $\tilde t c\chi^0$ coupling
$Y$ modulo the $b_i$ coefficients. For details see text. Here
$t_\beta\equiv \tan \beta$ and $a_u\equiv Av_u/\tilde m^2$.}
\label{tab:couplings}
\begin{center}
\begin{tabular}{l|cc} \hline\hline
\rule{0pt}{1.2em}%
%\label{tab:bqqq}
%\settabs 5 \columns
 & $\tilde t_L$ & $\tilde t_R$ \cr \hline
$\tilde H_u^0$\ & $\ \lambda_c\lambda_b^2
V_{cb}V_{tb}^*\sim4\times10^{-7}t_\beta^2\ $ &
$\ \lambda_c \lambda_t\lambda_b^2
V_{cb}V_{tb}^* \frac{A v_u}{\tilde m^2}\sim4\times10^{-7}t_\beta^2a_u\
$ \cr
$\tilde B$\ & $\ \sqrt{2} g^\prime Y_Q  \lambda_b^2V_{cb}V_{tb}^*\sim
3 \times 10^{-6}t_\beta^2$ &
$\ \sqrt{2} g^\prime Y_Q \lambda_t\lambda_b^2V_{cb}V_{tb}^* \frac{A
  v_u}{\tilde m^2} \sim  3 \times 10^{-6}t_\beta^2a_u$ \cr
$\tilde w^0$ & $\ \sqrt{2} g I_3 \lambda_b^2V_{cb}V_{tb}^*\sim
2\times10^{-5}t_\beta^2$ & $\ \sqrt{2} g I_3 \lambda_t \lambda_b^2
V_{cb} V_{tb}^*\frac{A v_u}{\tilde m^2} \sim
2\times10^{-5}t_\beta^2 a_u$ \cr
 \hline\hline
\end{tabular}
\end{center}
\end{table}

The leading couplings to the left-handed stop are induced by $(\tilde
m_{Q_L})_{23}$, whereas the ones to the right-handed stop component by
$(A_u)_{23}$. (The $c_1$ term gives only a subleading contribution.)
The higgsino-stop couplings receive an additional suppression from the
charm Yukawa.  The (hyper)charge assignments and the gauge coupling
suppress the stop-bino with respect to the stop-wino interaction.

Depending on the value of the soft parameters $A$ and the overall
squark mass scale $\tilde m$, the neutralino coupling to the $\tilde
t_R$ varies and can differ from the one to the $\tilde t_L$. In case
that $A v_u/\tilde m^2 \sim 1$, the couplings to $\tilde t_L$ and
$\tilde t_R$ are of the same size and, moreover, the light stop mass
eigenstate has comparable components of the two: $\theta_{\tilde t}
\sim m_t (A -\mu/\tan \beta)/\tilde m^2$. Such large $\tilde
t_L-\tilde t_R$ mixing is required in scenarios with a light stop mass
as low as ${\cal{O}}(100)$ GeV to lift the lightest Higgs mass above
the experimental limit, see, for instance, \cite{Essig:2007vq}. In
case that $A v_u/\tilde m^2 \ll 1$, the couplings of $\tilde t_R$ are
correspondingly smaller than those of $\tilde t_L$ and, furthermore,
the light stop could be dominantly $\tilde t_R$.

There can be further, model-dependent suppression of the $\tilde t
c \chi^0$ coupling $Y$ if $b_i$ and/or $c_i$, the coefficients of the
flavor changing squark mass-squared terms, are small.
For a generic MFV model, $b_i,c_i\lsim1$. In models where, at the
scale of supersymmetry breaking mediation the soft terms are universal
($b_i=c_i=0$), they are nevertheless generated at lower scales by
renormalization group evolution (RGE).
The $b_i$ coefficients are generated at one loop, while $c_1$ is
generated at two loops (see, for example, Ref. \cite{Martin:1993zk}).
It thus makes sense to consider as lower bounds $b_i\gsim(1/16\pi^2)
\times logs$ and $c_1\gsim(1/16\pi^2)^2 \times logs$.  We term
$b_i,c_1\sim1$ as ``weak MFV'' and $b_i\sim10^{-2}$, $c_1\sim10^{-4}$
as ``strong MFV''. Weak MFV can be realized with, for example, Yukawa
deflected gauge mediation \cite{Chacko:2001km,Chacko:2002et}, while
strong MFV can be realized with, for example, standard low energy
gauge mediation \cite{Dine:1995ag}. An experimental determination of
the $b_i,c_i$ is possible from the stop FCNC coupling $Y$ once
MFV has been established and the stop and neutralino decomposition and
$\tan \beta$ are known.

To summarize, the range of $Y$ covered in MFV models is given as
\beq\label{ymfv}
 10^{-10} \lesssim  Y \lesssim 10^{-4} ~~(\rm{strong ~MFV}), ~~~~~
 10^{-8} \lesssim  Y \lesssim 10^{-2} ~~(\rm{weak ~MFV}).
\eeq
The upper bounds are reached with large $\tan \beta \sim 30$ whereas for
the lower bounds we assumed $\tilde t_L-\tilde t_R$ mixing above the
percent level.

%%%%%%%%%%%%%%
\section{Measuring flavor changing decay rates at the LHC}
\label{sec:lhc}
If the flavor structure of the squark mass matrices is minimally
flavor violating, and if the decays of the lightest stop into third
generation quarks are kinematically forbidden, the stop will be
surprisingly long-lived. In particular, its lifetime may be long
enough that its decay may give a signature of a secondary vertex at
the ATLAS and CMS experiments.

The precise minimal decay length that will allow a measurement of the
lifetime depends on various details of the detector (see, for example,
the discussion in \cite{Ragusa:2007zz}) and on the typical boost of
the stop squarks. We assume here that $\beta\gamma$ is of order one,
and take the minimal lifetime to be measured via secondary vertex as
0.3 ps, corresponding to a decay length of 0.1 mm.
From Eq.~(\ref{lifetime}) we learn that the lifetime will be long
enough for a measurement if
\beq \label{eq:condition}
Y (\Delta m/M)\lsim5\times10^{-7}.
\eeq

When we estimate the lifetime of the light stop, we have to take into
account the decomposition of the stop and of the neutralino,
and the flavor suppression factors of Table \ref{tab:couplings}. In
addition, within our scenario, the phase space factor provides further
suppression, $\Delta m/M\sim0.03$, and there can be weak or strong MFV
suppression in the $b_i,c_i$ coefficients, which have been discussed
in Section \ref{sec:mfv}.

We remark that there are also charged current contributions to $\tilde
t \to c \chi^0$ decays induced at one loop \cite{Hikasa:1987db}.  MFV
enforces that their flavor structure is the same as that of the
corresponding $Y$.  The dominant, logarithmic part of the loops stems
from the soft mass counter term and induces the same decay amplitude
as the pure RGE contribution \cite{Hikasa:1987db}. It is hence
included in $Y$ with the $b_i$ taken at the low scale.  The remaining,
non-logarithmic corrections from the loops to the relation between $Y$
and the decay rate Eq.~(\ref{decrat}) are subleading and can be
neglected for this study.

\begin{table}[t]
\caption{The numerical size of $Y(\Delta m/M)$ with $Y$ taken from
  Table \ref{tab:couplings}, $\Delta m/M\sim0.03$, $\tan\beta\sim3$,
$A v_u/\tilde m^2 \sim 1$
  and $b_i\sim1$ (weak MFV) and $b_i \sim 10^{-2}$ (strong MFV).}
\label{tab:weak}
\begin{center}
\begin{tabular}{l|cc} \hline\hline
\rule{0pt}{1.2em}%
%\settabs 5 \columns
 &  weak MFV & strong MFV \cr \hline
$\tilde H_u^0$\ & $1 \times 10^{-7}\ $ & $1 \times 10^{-9}\ $ \cr
$\tilde B$\ & $9 \times10^{-7}$\ &\ $9\times10^{-9}$\ \cr
$\tilde w^0$\ & $5 \times 10^{-6}$ & $5 \times 10^{-8}$ \cr
 \hline\hline
\end{tabular}
\end{center}
\end{table}

Our estimate of $Y(\Delta m/M)$ for both MFV cases are given in Table
\ref{tab:weak}. Here, we assume $A v_u/\tilde m^2 \sim 1$, hence the
couplings to $\tilde t_L$ and $\tilde t_R$ are of the same size and
not given separately. Comparing this table with
Eq.~(\ref{eq:condition}), we conclude that it will be possible to
measure the lifetime of the light stop for a rather large part of the
parameter space. In particular, within our scenario and working
assumptions, a measurement will be possible for low values of $\tan
\beta \sim 3$ in the weak MFV scenario if the light neutralino is
predominantly the higgsino or bino ($|V_{1w}|\lsim0.1$) or, for any
decomposition of the neutralino, if $Av_u/\tilde m^2\lsim0.1$ (and
correspondingly $\sin\theta_{\tilde t}\lsim0.1$). As concerns strong
MFV, we find that the stop lifetime is longer than 30 ps in the entire
parameter space.

Keeping the masses fixed, the $\tan \beta$ dependence of our findings
is dominated by the two powers of the bottom Yukawa in $Y$, see Table
\ref{tab:couplings}. The $\tilde t \to c \chi^0 $ decay rate hence
exhibits a steep $\tan \beta$ dependence, $\Gamma \propto \tan^4
\beta$, and the region in parameter space with visible secondary
vertex gets constrained towards larger values of $\tan \beta$.  For
example, for $\tan \beta \sim 10$, Eq.~(\ref{eq:condition}) requires
in the weak MFV scenario the stop to be right-handed with an $\tilde
t_L$ admixture of at most a few percent. In scenarios with strong MFV
suppression the lifetime can still be measured in the whole region
given a higgsino-type neutralino or a mostly bino gaugino
($|V_{1w}|\lsim0.9$) or simply $\sin\theta_{\tilde t}\lsim0.9$.

Light stops have been searched for already at colliders assuming that
the dominant decay mode is $\tilde t \to c \chi^0$, that is, in
missing energy plus jet signatures~\cite{Abazov:2006wb,overview}.  A
discovery in this channel will determine the mass, but not the stop
flavor couplings, which can be extracted from analyzing the stop decay
length.

%%%%%%%%%%%%%%%%%%%%%
\section{$m_{\tilde t}-m_{\chi^0}>m_b$ and four body decays}
\label{sec:4body}
Identifying a secondary vertex from a charm jet with energy of the
order $\Delta m \leq m_b$ times a boost factor, see Eq.~(\ref{degen}),
is experimentally challenging. Therefore, and also to understand the
general situation in supersymmetric models, we would like to
investigate the possibility of relaxing the constraint on the
stop-neutralino splitting. In other words, we still consider a
scenario where the light stop is the NLSP, but with larger mass
splitting, $m_{\tilde t}-m_{\chi^0}>m_b$.

Our proposal to measure MFV couplings is based on the dominance of
$\tilde t \to c \chi^0$ decays and as long as this is true, $\Delta m$
could be larger. Since the light stop is the NLSP, decays such as
$\tilde t \to b \chi^+$ are forbidden. The four body decays $\tilde t
\to b \chi^0 l \nu$ \cite{Hikasa:1987db} are, however, kinematically
open. Several diagrams contribute at tree level $\sim V_{tb}^*$, with
the potentially dangerous ones containing charginos and the $W$-boson
or sleptons \cite{Boehm:1999tr}.

We give here a rough estimate of these contributions to the four body
decay rate. The matrix elements squared of the leading diagrams of
$\tilde t \to b \chi^0 l \nu$ decays go with the third power of light
fermion ($b,l,\nu$) momenta.  Furthermore, from phase space we get
five powers of light momenta\footnote{This can be shown by 
performing the phase space integration analytically assuming a flat 
matrix element. We thank Stephen Martin for clarifying this point.}.
This suggests that $\Gamma_{4-{\rm
    body}} \sim (\delta m)^8/(M m_W^4 m_{\chi^+}^2)$ or 
$(\delta m)^8/(M m_{\tilde l}^4 m_{\chi^+}^2)$ where we assumed $m_{\tilde t,
  \chi^0} \ll m_{\chi^+, \tilde l}$ and $\delta m = \Delta m -m_b$
is the available kinetic energy.  We obtain for the (leading)
$W$-contribution:
\beq \label{eq:ratio}
\frac{\Gamma(\tilde t\to b\chi^0 l\nu)}{\Gamma(\tilde t\to c\chi^0)}
\approx\frac{g^6|V_{tb}|^2}{2 (4 \pi)^4}\frac{(\Delta m-m_b)^8}
{[Y \Delta m]^2  m_W^4 m_{\chi^+}^2}.
\eeq
Despite the substantial mass and phase space suppression of this
ratio, numerically it turns out that, for $m_{\chi^+}$ below 500
GeV, $\Delta m$ can only be lifted by $O(10)$ GeV above the bottom mass
without invalidating our assumptions. This is caused by the smallness
of the coupling $Y$ in the denominator, for which we require to yield
a macroscopic decay length, {\it i.e.}, Eq.~(\ref{eq:condition}).

\begin{figure}
\includegraphics[scale=1.0]{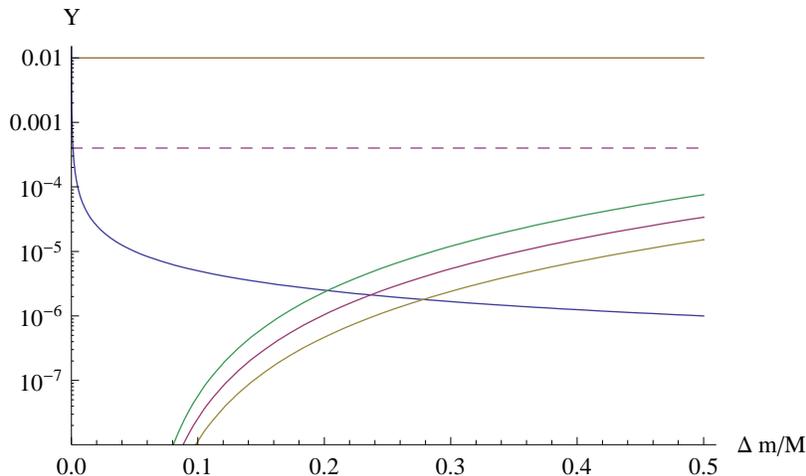}
\caption{\label{fig:Ydm}Interesting regions in the $Y-\Delta m/M$
  plane: (i) The single curved line separates the regions where a
  secondary vertex appears (below) or does not appear (above); (ii)
  The triplet of curved lines distinguishes the region where the stop
  two-body decay dominates (left and above) from the region where the
  four-body decays dominate (below and right); (iii) The horizontal dashed
  line marks the lower bound on the region accessible to models with
  alignment; (iv) The horizontal solid line marks the lower bound if 
  the up squark flavor parameters are anarchical. The plot is shown
  for fixed $m_{\tilde t}=100$ GeV and  $m_{\chi^+}=500$ GeV. For 
  details see text.}
\end{figure}

The situation is schematically depicted in Fig. \ref{fig:Ydm}. We
distinguish various interesting regions in the $Y-\Delta m/M$ plane,
shown for fixed $m_{\tilde t}=100$ GeV and $m_{\chi^+}=500$ GeV:
\begin{itemize}
\item[{\it (i)}] Below the single curved (blue) line, the stop
  lifetime is long enough for a secondary vertex to appear, that is,
  Eq.~(\ref{eq:condition}) is fulfilled.
\item[{\it (ii)}] Above and to the left of the triple lines, the
  $\tilde t\to c\chi^0$ decay dominates. More precisely, the ratio of
  Eq.~(\ref{eq:ratio}) is smaller than 5,1,1/5, for the lower, middle
  and upper line, with the spread modeling the uncertainty of our
  estimate for the $\tilde t \to b \chi^0 l \nu $ decay rate.
\item[{\it (iii)}] Above the horizontal dashed line at
  $Y=\lambda_c V_{cb}=4\times10^{-4}$ is the region accessible to
  alignment models (see Section \ref{sec:nonmfv} for details).
\item[{\it (iv)}] Above the horizontal solid line at $Y=0.01=\lambda_c$
  is the region accessible with 'squark flavor anarchy', {\it i.e.} no
  special structure in the relevant soft supersymmetry breaking terms
  (see Section \ref{sec:nonmfv}). 
\end{itemize}
Our stop search strategy works in the lower left corner of the
$Y-\Delta m/M$ plane.

%%%%%%%%%%%%%%%%%%%%%%%%%%%%

With the requisite replacements, it follows also from
Eq.~(\ref{eq:ratio}) that the CKM suppressed modes $\tilde t\to
s\chi^0 l\nu$, which are not excluded by the mass constraint
Eq.~(\ref{degen}), are not competitive with $\tilde t \to c \chi^0$
decays for $\Delta m$ below ${\cal{O}}(10-20)$ GeV.

%%%%%%%%%%%%%%%%%%
\section{A light chargino}
\label{sec:chargino}
It is interesting to consider the case where the light stop $\tilde t$
is the NNLSP, with the lighter chargino $\chi^+$ being the NLSP:
\beq
m_{\tilde  t}> m_{\chi^+}>m_{\chi^0},
\eeq
which we assume in addition to the condition (\ref{degen}).
Then, besides the decay mode (\ref{domdec}), the stop can decay through
\beq\label{sdomdec}
\tilde t\to\chi^++s.
\eeq
We consider an arbitrary decomposition of the light chargino,
\beq
\chi^+ (\equiv \chi^+_1)=\cos\theta_+\tilde w^++\sin\theta_+\tilde H^+,
\eeq
where $\tilde H^+=\sin\beta \tilde H_d^{*+}+\cos\beta \tilde H_u^+$.

A crucial point here is that, unlike the case of a final neutralino,
we now have flavor changing couplings even in the supersymmetric
limit. This is in correspondence to the fact that, within the
Standard Model, there are flavor changing charged current
interactions but (at tree level) no flavor changing neutral
current interactions. We have the following relevant couplings:
\begin{itemize}
\item The $\tilde t_L-s_L-\tilde w^+$ coupling is related via
supersymmetry to the SM $t_L-s_L-W^+$ coupling.
\item The $\tilde t_L-s_R-\tilde H^+$ coupling comes from
  the down Yukawa coupling and quark mixing.
\item  The $\tilde t_R-s_R-\tilde w^+$ coupling vanishes.
\item The $\tilde t_R-s_L-\tilde H^+$ coupling is given by the up
  Yukawa coupling and quark flavor mixing.
\end{itemize}
The situation is summarized in Table \ref{tab:coupc}. In addition
to the previously used parameters, we use $|V_{ts}|\sim0.04$ and
$\lambda_s\sim 5\times10^{-4}\tan\beta$.

\begin{table}[t]
\caption{Flavor structure  and numerical size of the supersymmetric
  contributions to the $\tilde t s\chi^+$ coupling. Here
  $t_\beta\equiv\tan\beta$.}
\label{tab:coupc}
\begin{center}
\begin{tabular}{c|cc} \hline\hline
\rule{0pt}{1.2em}%
%\label{tab:bqqq}
%\settabs 5 \columns
 & $\tilde t_L$ & $\tilde t_R$ \cr \hline
$\tilde H^+$ & $\ \sin\beta\lambda_s V_{ts}^*\sim2\times10^{-5}t_\beta$ &
 $\ \cos\beta\lambda_t V_{ts}^*\sim4\times10^{-2}/t_\beta\ $ \cr
$\tilde w^+$ & $g V_{ts}^*\sim3\times10^{-2}$ &
$0$   \cr
 \hline\hline
\end{tabular}
\end{center}
\end{table}
We learn that the flavor suppression of the chargino modes is not
strong enough to induce a secondary vertex.  Even assuming a pure
gaugino and a $\tilde t_L$ admixture as small as few $\times 10^{-4}$
with $\Delta m/M \sim 0.03$ violates the condition
Eq.~(\ref{eq:condition}).  It will therefore be difficult to establish
the CKM suppression of the stop decays into lighter generations if the
chargino is lighter than the stop.

It is amusing to note that if the stop-chargino degeneracy is strong
enough that even the decays to final states with strangeness are
kinematically forbidden, then the decay rate is further suppressed by
the smaller phase space $\Delta m/M\sim10^{-3}$, by the smaller CKM element,
$|V_{td}|\sim0.2|V_{ts}|$, and, where relevant, by a smaller Yukawa
coupling, $\lambda_d\sim0.05\lambda_s$. Still, in most of the
parameter space, the decay length will be too short to be measurable.

In any case, we should emphasize that the CKM suppression of the
charged current $\tilde ts\chi^+$ coupling as shown in Table
\ref{tab:coupc} is generic in supersymmetry. It is a consequence of
supersymmetry, and is not related to the question of whether the
mediation of supersymmetry breaking is MFV or not. However, beyond
MFV, squark loops can alter charged current couplings significantly
from their tree level values, which can be used to signal the
breakdown of MFV, see \cite{Dittmaier:2007uw} for an LHC example.

%%%%%%%%%%%%%%%%%%
\section{Third generation flavor mixing without MFV}
\label{sec:nonmfv}
The significance of testing the MFV hypothesis can be appreciated by
investigating models without MFV. One should ask, first, whether there
are natural and viable models of supersymmetry breaking that do not
implement the MFV principle and, second, whether their predictions for
the flavor changing couplings are significantly different from those
of MFV models.

In the case of `anarchical' soft supersymmetry breaking parameters,
there is no CKM suppression and the flavor changing $\tilde t c \chi^0$ 
vertex is
generically $Y \sim \sqrt{2} g I_3, \sqrt{2} g^\prime Y_Q, \lambda_c$
for the wino, bino and higgsino, respectively. However, applying
experimental constraints from FCNCs excludes such generic models. A
better framework would be one with a natural mechanism to
suppress flavor changing couplings. An example of such a framework is
that of alignment \cite{Nir:1993mx,Leurer:1993gy}.

Models of alignment are based on an Abelian horizontal symmetry that
is broken by small parameters \cite{Froggatt:1978nt}. In the simplest
version there is a single $U(1)_H$ which is broken by a single spurion
$\epsilon$ of charge $H=-1$. Then, the charges of the various
superfields are determined by the measured quark parameters:
\beq
|V_{ij}|\sim\epsilon^{H(Q_{Li})-H(Q_{Lj})}\ (j>i),\ \ \ \ 
\lambda_{u_i}\sim\epsilon^{H(Q_{Li})+H(\bar U_{Ri})}.
\eeq
The same charges determine also the parametric suppression of the soft
supersymmetry breaking parameters, leading to the following order of
magnitude relations:
\beqa
(\tilde m^2_{Q_L})_{23}&\sim&\tilde m^2 V_{cb},\no\\
(\tilde m^2_{U_R})_{23}&\sim&\tilde m^2 \lambda_c/(\lambda_t V_{cb}),\no\\
(A_u)_{23}&\sim&A\lambda_tV_{cb},\no\\
(A_u)_{32}&\sim&A\lambda_c/V_{cb}.
\label{eq:nonMFV23}
\eeqa
We now use the procedure described in Section \ref{sec:mfv},
inserting, however, the order of magnitude estimates of
Eq.~(\ref{eq:nonMFV23}) instead of those of Eq.~(\ref{eq:MFV23}). 
This leads to the suppression factors presented in Table
\ref{tab:coupn}. We denote by $Y_U=2/3$ the hypercharge of the (s)charm
singlet, other parameters are as in Section \ref{sec:mfv}.

%%%%%%%%%%%%%%%%%%%%%%%%%%%%%
\begin{table}[t]
\caption{
Flavor structure and numerical size of the $\tilde t c\chi^0$ coupling
$Y$ in naive alignment models. Here $a_u\equiv Av_u/\tilde m^2$.}
\label{tab:coupn}
\begin{center}
\begin{tabular}{l|cc} \hline\hline
\rule{0pt}{1.2em}%
%\label{tab:bqqq}
%\settabs 5 \columns
 & $\tilde t_L$ & $\tilde t_R$ \cr \hline
$\tilde H_u^0$\ & $\ {\rm max}(\frac{\lambda_c^2}{V_{cb}}\frac{A
  v_u}{\tilde m^2},\lambda_cV_{cb})
\sim{\rm max}(2\times10^{-3}a_u,4\times10^{-4})\ $ &
$\ \frac{\lambda_c^2}{\lambda_t V_{cb}}\sim2\times10^{-3}\ $ \cr
$\tilde B$\ & $\ \sqrt{2} g^\prime{\rm max}(Y_U\frac{\lambda_c}{V_{cb}}
\frac{A v_u}{\tilde m^2},Y_QV_{cb})\sim {\rm max}(0.08a_u,3 \times 10^{-3})$ &
$\ \sqrt{2} g^\prime Y_U \frac{\lambda_c}{\lambda_tV_{cb}} \sim 0.08$ \cr
$\tilde w^0$ & $\ \sqrt{2} g I_3 V_{cb}\sim 0.02$ &
$\ \sqrt{2} g I_3 \lambda_t V_{cb} \frac{A v_u}{\tilde m^2}\sim 0.02 a_u$ \cr
 \hline\hline
\end{tabular}
\end{center}
\end{table}
%%%%%%%%%%%%%%%%%%%%%%%%%%%%%%%%%%%

For $\tan\beta\sim3$ and $a_u\sim1$, the alignment couplings are
larger by two to three orders of magnitude than the (weak) MFV ones given in
Table \ref{tab:couplings} for $b_i \sim 1$. In the
very large $\tan\beta$ limit, where the $\lambda_b^2$ suppression of
the MFV flavor changing couplings is ineffective, the two models can
give comparable couplings. In general, for $a_u\gsim0.01$, alignment
models span the following range:
\beq\label{ynonmfv}
10^{-4} \lesssim  Y \lesssim 10^{-1} ~~(\rm{Alignment}),
\eeq
to be compared with the MFV range in Eq.~(\ref{ymfv}). The most
important difference here is that in alignment models the resulting 
stop lifetime is short, $\tau_{\tilde t}^{\rm alignment} \sim (10^{-20}-
10^{-15})$ s, and hence, alignment models are not expected
to give a secondary vertex signal.

%%%%%%%%%%%%%%%%%%
\section{Conclusions}
\label{sec:con}
The question of whether the mechanism that mediates supersymmetry
breaking is MFV is important and provides a window to scales well
beyond the direct reach of the LHC. While it may be easy to exclude
MFV if it is violated in a strong way, it will be a much more
challenging task to experimentally establish MFV in case that it
holds. One model-independent prediction of MFV is a high degeneracy between
the first two squark generations, below a GeV \cite{Hiller:2001qg}.
Its measurement at the LHC will, like the CKM and Yukawa suppression of
the flavor mixing, be most likely impossible.

We point out that, under a certain set of circumstances, measuring the
mixing within MFV models might be possible after all. This set of conditions 
requires that the stop is the NLSP, and that its splitting from the
neutralino-LSP is not much bigger than the mass of the bottom quark. Then the
light stop decays predominantly into second (or first) generation quarks.
Furthermore, the CKM suppression, the Yukawa suppression, and the phase-space
suppression combine to make the lifetime of the stop unusually long.
In fact, it is long enough that the decay might occur with a secondary
vertex. This is the crucial ingredient that may provide ATLAS and CMS
with a way to measure the lifetime, and by that provide information on
the size of the flavor changing couplings related to the breaking of
supersymmetry. 

The flavor suppression that is required to provide (in combination
with the accidental stop-neutralino degeneracy) a stop lifetime that
is long enough to generate a secondary vertex is quite unique to MFV
models. Observing such a secondary vertex, even without a precise
determination of the lifetime, would lend strong support to the MFV
principle.

Light stops and MFV are features, for instance, of models with
hypercharged anomaly mediation \cite{Dermisek:2007qi}. Here the light
stop is mostly left-handed and the neutralino LSP is mostly wino, such
that a stop lifetime measurement would work up to moderate values of
$\tan \beta$. Note also that our generic requirement of a small mass
gap between the lightest stop and the lightest neutralino supports
efficient coannihilation in the neutralino relic density calculation
\cite{Boehm:1999bj,Balazs:2004bu}.

We conclude that a flavor program in ATLAS and CMS can be of
unique capability in addressing the flavor puzzles
\cite{Dittmaier:2007uw,Feng:2007ke,Nomura:2007ap}.

\acknowledgments
We thank Michael Peskin for pointing out to us that third generation
squark decays into third generation quarks might be kinematically
forbidden. This work was supported by a grant from the G.I.F., the
German--Israeli Foundation for Scientific Research and Development.
The work of YN is supported in part by the United States-Israel
Binational Science Foundation (BSF), by the Israel Science Foundation
(ISF), and the Minerva Foundation.

%%%%%%%%%%%%%%%%%%%%%%%%%%%%%%%%%%%%%%%%%%%%%%%%%%%%%%

%%%%%%%%%%%%%%%%%%%%%%%%%%%%%%%%%%%%%%%%%%%%%%%%%%%%%%

\end{document}